# High Dynamic Range Externally Time-gated Photon Counting Optical Time-domain Reflectometry

Bin Li[2], Guangwei Deng[1,†], Ruiming Zhang[1], Zhonghua Ou[3], Heng Zhou[2], Yun Ling[2], Yunxiang Wang[3], You Wang[1,4], Kun Qiu[2], Haizhi Song[1,4,‡], and Qiang Zhou[1,3,5*]

[1]*Institute of Fundamental and Frontier Sciences, University of Electronic Science and Technology of China, Chengdu 610054, PR China*
[2]*School of Information and Communication Engineering, University of Electronic Science and Technology of China, Chengdu 610054, PR China*
[3]*School of Optoelectronic Science and Engineering, University of Electronic Science and Technology of China, Chengdu 610054, PR China*
[4]*Southwest Institute of Technical Physics, Chengdu 610041, PR China*
[5]*CAS Key Laboratory of Quantum Information, University of Science and Technology of China, Hefei 230026, PR China*

*Abstract*—Single photon detector (SPD) has a maximum count rate due to its dead time, which results in that the dynamic range of photon counting optical time-domain reflectometry (PC-OTDR) decreases with the length of monitored fiber. To further improve the dynamic range of PC-OTDR, we propose and demonstrate an externally time-gated scheme. The externally time-gated scheme is realized by using a high-speed optical switch, i.e. a Mach-Zehnder interferometer, to modulate the back-propagation optical signal, and to allow that only a certain segment of the fiber is monitored by the SPD. The feasibility of proposed scheme is first examined with theoretical analysis and simulation; then we experimentally demonstrate it with our experimental PC-OTDR testbed operating at 800 nm wavelength band. In our studies, a dynamic range of 30.0 dB is achieved in a 70 meters long PC-OTDR system with 50 ns external gates, corresponding to an improvement of 11.0 dB in dynamic range comparing with no gating operation. Furthermore, with the improved dynamic range, a successful identification of a 0.37 dB loss event is detected with 30-seconds accumulation, which could not be identified without gating operation. Our scheme paves an avenue for developing PC-OTDR systems with high dynamic range.

*Index Terms*—Fiber testing, gating operation, optical time-domain reflectometry, photon counting.

## I. INTRODUCTION

OPTICAL time-domain reflectometry (OTDR) is a representative distributed fiber optic sensor. Since its first demonstration in 1976 [1], it has been successfully employed in many fields such as optical fiber communication [2], high-voltage transmission lines [3], oil and gas pipelines [4], etc. Most of commercially available OTDR systems are based on linear photon detectors, such as PIN or avalanche photodiodes (APDs). Conventional OTDR has great advantages in long distance measurement and has achieved a dynamic range of up to ~50.0 dB with spatial resolutions of around one kilometer [5]. However, the improvement of its spatial resolution is still a challenge. This is due to the limitation of detection bandwidth of photodetectors. On the other hand, the intensity of the back-propagation signal decreases with the width of laser pulse. Therefore, it requires a photodetector with high sensitivity and large bandwidth simultaneously. However, photodetectors under linear regime with large bandwidth and high-sensitivity are too difficult to develop, thus preventing the development of the conventional OTDR system with high spatial resolution and dynamic range [6]-[8].

In 1980, photon-counting OTDR (PC-OTDR) based on the Geiger-mode single photon detector (SPD) was proposed and demonstrated [9]. Compared with the conventional OTDR, PC-OTDR can offer advantages of higher spatial resolution and larger dynamic range [10] [11]. Especially, with the development of single photon detection technology, Q. Zhao *et al.* obtained a spatial resolution of 4 mm with superconducting nanowire single-photon detectors (SNSPDs) [12], J. Hu *et al.* exhibited a dynamic range of 22 dB m with 6.0 cm spatial resolution at the end of 2 km standard single-mode fiber [13], and G.-L. Shentu *et al.* achieved a dynamic range of 42.19 dB with a spatial resolution of 10 cm with an ultra-low noise up-conversion SPD [14]. Further improvement of the dynamic range of PC-OTDR is limited by parameters of SPD. One key parameter is the so-called dead time, which is the minimum time needed to recover the SPD after a detection event. Therefore, the dead time sets a maximum count rate for SPD [15]. For a certain PC-OTDR system, the maximum photon counting rate, combining with the noise, i.e. the dark count rate of the SPD, gives the upper limit of the dynamic range for PC-OTDR system. That is to say, the longer the measured fiber, the fewer number of photons accumulated from fiber per meter - the number of photons is fixed by the maximum count rate of the SPD, thus leading to the reduction of the dynamic range of PC-OTDR. In other words, the dynamic range for a given PC-OTDR system could not be improved by arbitrarily increasing the back-propagation signal, due to the SPD has a saturation count rate.

We propose and demonstrate a novel externally time-gated PC-OTDR scheme. Our scheme uses a high-speed optical switch (OSW), i.e. Mach-Zehnder interferometer (MZI) based intensity modulator, to modulate the back-propagation signal, and to allow that only a certain part of the back-propagation signal - defined by the gate - is detected by the following SPD. Alternatively, one can internally gate the SPD by using an electronic gate, such that the detector is active only during the gate period [16]. Compared with the internally time-gated method, our externally time-gated scheme can eliminate the counting spikes on the rising edge of the internal gate - Appendix A for details. In this paper, we present the scheme of externally time-gated PC-OTDR. Then, a theoretical model is developed, which helps us to analyze the performance of our scheme. Finally, we examine our scheme with experimental PC-OTDR testbed. Our experimental results show that with 50 ns external gates a dynamic range of 30.0 dB can be achieved for a 70 meters long PC-OTDR test, showing an improvement of 11.0 dB in dynamic range comparing with no gating operation. Furthermore, with the improved dynamic range, a successful identification of a 0.37 dB loss event is detected with 30-seconds accumulation, which is not be identified without gating operation in the experiment. Our scheme paves an avenue for achieving a high dynamic PC-OTDR system.

†gwdeng@uestc.edu.cn;
‡hzsong1209@163.com;
*zhouqiang@uestc.edu.cn

## II. THEORETICAL ANALYSIS FOR EXTERNALLY TIME-GATED PC-OTDR SYSTEM

### A. Basic Operation

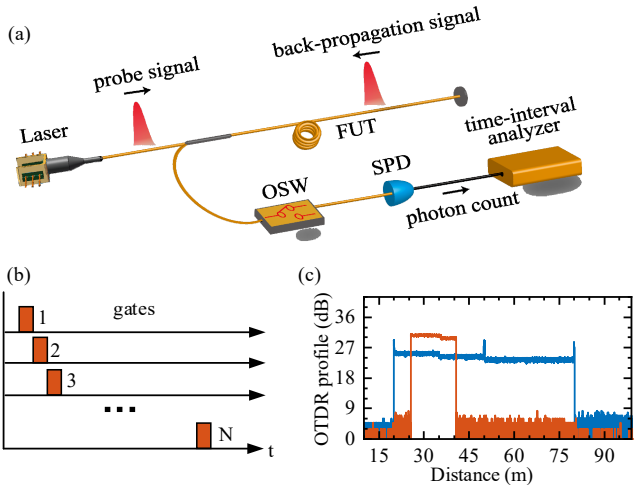

Fig. 1. (a) Schematic diagram of externally time-gated PC-OTDR. (b) Gated signals provided to optical switch (OSW). (c) Simulated OTDR traces of 60 m fiber link using a laser pulse width of 1 ns. The blue curve represents the result of OTDR in standard configuration in which no gated signal is used. The peak power of the probe signal is 10 µW. The result of externally time-gated scheme is represented by the orange curve using an OSW with an extinction ratio of 30.0 dB and a gate width of 15 ns. In order to keep the same count rate from the single photon detector (SPD), the peak power of the probe signal is increased from 10 µW to 39 µW in the simulation.

As shown in Fig. 1, the idea of externally time-gated PC-OTDR is that we divide the OTDR trace into $N$ segments by using an OSW. Each segment is defined by the external gate which controls the OSW. After obtaining the $N$ segments of OTDR trace, we seam them together in sequence, and recover the whole trace. By doing so, for each segment of OTDR trace, we can increase the corresponding back-propagation signal to the saturation count rate of the SPD, by improving the intensity of probe signal, thus increasing the signal-level for the PC-OTDR trace. While the noise floor does not change, and is determined by the dark count rate of the SPD. Thus, the dynamic range of the externally time-gated PC-OTDR is improved with the external gating operation. Note that strong Fresnel reflections may take place at connection points along the fiber link, which may saturate the SPD at those places. Fortunately, the Fresnel reflection only happens at a single point, which means that with the proposed externally time-gated scheme these Fresnel reflection points can be skipped by precisely controlling the timing of the external gate, then leaving only the Raleigh scattering is measured along the OTDR trace. Furthermore, the location of all Fresnel reflection points can be automatically located by event location algorithms, such as wavelet transform [17], trend filter [18], etc.

### B. Theoretical Model

According to [19], we develop a theoretical model to analyze the feasibility of the proposed scheme. We consider that the fiber under test (FUT) is composed of $N_b$ scattering units. For the case without external gating operation, let $P_{BS}(i)$ be the backscatter optical power from each scattering unit. The total photon count is then given by the sum of each scattering unit,

$$C_1 = \sum_{i=1}^{N_b} \frac{P_{BS}(i) \times t_{total}}{h\nu}. \quad (1)$$

Where $h\nu$ is the energy per backscattered photon, $t_{total}$ is the measurement time. On the other hand, when we use externally time-gated technique and divide the PC-OTDR trace into $n$ segments by gated signals, each segment contains $m$ (floor of $N_b / n$) scattering units. Due to the change of input power, the backscatter power for each scattering unit within the gate becomes $P'_{BS}(i)$ and that outside of the gate is $P'_{BS}(i) / \gamma$, where $\gamma = 10\exp(E_R / 10)$, $E_R$ is the extinction ratio of the OSW. Hence, for the $k$-th gated signal, the number of photons is calculated as

$$C(k) = \sum_{i=1}^{(k-1)\times m} \frac{1}{\gamma} \times \frac{P'_{BS}(i) \times t_{each}}{h\nu} + \sum_{i=(k-1)\times m+1}^{k\times m} \frac{P'_{BS}(i) \times t_{each}}{h\nu} + \sum_{i=k\times m+1}^{N_b} \frac{1}{\gamma} \times \frac{P'_{BS}(i) \times t_{each}}{h\nu}, \quad (2)$$

where $t_{each}$ is the time required for each gated test and $t_{each} = t_{total} / n$. The three terms in (2) in order are the count generated by the scattering units before, within and after the gate, respectively. Finally, after recovering the whole trace, we infer the total photon count of our scheme,

$$C_2 = \sum_{k=1}^{n} C(k). \quad (3)$$

To simplify the analysis, we assume $P_{BS}(i) \cong P_{BS}(1)$ and $P'_{BS}(i) \cong P'_{BS}(1)$, which is reasonable when the length of FUT is less than 1 km. And the total photon counts for both cases of with and without gating operation are the same, i.e. $C_1 = C_2$, which are determined by the dead time of SPD. Therefore, we can obtain the improvement of dynamic range for the case with external gates [19],

$$\Delta D_R = 10\log(\frac{P'_{BS}(1)}{N_{EP0}}) - 10\log(\frac{P_{BS}(1)}{N_{EP0}}),  \quad (4)$$
$$= 10\log(\frac{\gamma \times N_b}{\gamma \times m + (N_b - m)})$$

where $N_{EP0}$ is the minimal detectable power of SPD, $\Delta D_R$ is the improvement of dynamic range. From (4), one can see that the dynamic range is improved by using the externally time-gated schematic, and increases with the increase of extinction ratio and the decrease of gate width (for details, see Appendix B). If $\gamma \gg N_b/m$, (4) can be further simplified as,

$$\Delta D_R \cong 10\log\left(\frac{N_b}{m}\right). \quad (5)$$

This approximation is valid for OSW with high extinction ratio ($E_R > 30.0$ dB) and less segments. For example, let assume $E_R = 50.0$ dB and $N_b/m = 10$, then the externally time-gated PC-OTDR achieves a 10.0 dB improvement in dynamic range. Fig. 1(c) shows the simulated OTDR traces of 60 m fiber link using a laser pulse width of 1 ns. The blue curve represents the simulated result of OTDR in standard configuration in which no gated signal is used. The peak power of the probe signal is 10 μW. The result of externally time-gated scheme is represented by the orange curve using an OSW with an extinction ratio of 30.0 dB and a gate width of 15 ns. In order to keep the same count rate from the SPD, the peak power of the probe signal is increased from 10 μW to 39 μW in the simulation. An improvement of a dynamic range of 6.02 dB is obtained, which corresponds to (5) with $N_b/m = 4$.

### III. EXPERIMENTAL DEMONSTRATION AND RESULTS

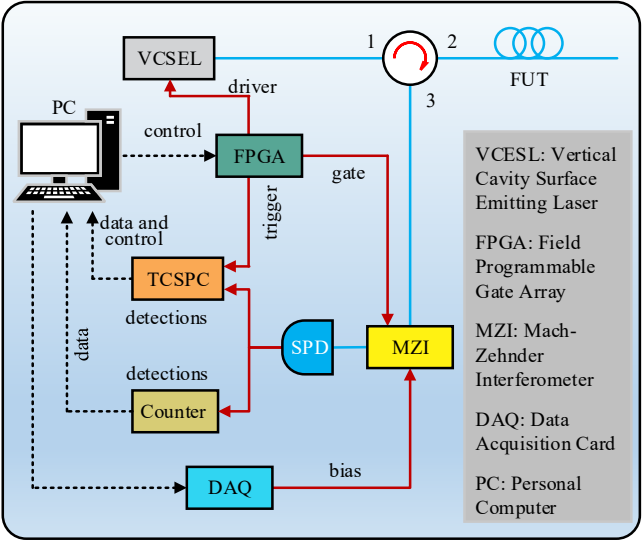

Fig. 2. Schematic of the externally time-gated PC-OTDR for fiber testing.

We experimentally demonstrate our externally time-gated scheme based on our PC-OTDR testbed at 850 nm [20]. Fig. 2 shows the detailed setups of our experiment. Laser pulses at 850 nm output from a vertical cavity surface-emitting laser (VCSEL) driven by the field programmable gate array (FPGA). The generated laser pulses are launched into the FUT through an optical circulator. The laser is set at a repetition rate of 1 MHz and at a pulse width of less than 1 ns, which is equivalent to a test range of 100 m and a spatial resolution of less than 10 cm. The back-propagation signal is coupled into the third port of a circulator and then feed into an MZI, which is the high-speed OSW in our experiment. The FPGA board creates the electronic gated signals, i.e. the modulation signals for MZI. The insertion loss of MZI is 1.5 dB, and its extinction ratio is 29.0 dB. Then, the modulated back-propagation signal is detected by the SPD (Excelitas, SPCM-AQRH-14). The dark count rate of the SPD is about 100 Hz while the detection efficiency is 45%, which gives a noise equivalent power (NEP) of about $7.35\times10^{-18}$ W/$\sqrt{\text{Hz}}$ [19]. The electronic signal output from SPD is split into two channels by using a T-connector. One feeds to the time correlated single photon counting (TCSPC, TimeHarp 260) board, and the other to a counter (Agilent, 53131A). The TCSPC board, working in histogram mode with a time bin width of 0.1 ns, is triggered by signals from FPGA. Finally, the data is transmitted to a computer through the peripheral component interconnect express (PCIE) interface for processing and analysis. It is worth noting that the bias point of MZI drifts with its temperature. To stabilize its bias point, a data acquisition card (DAQ, Texas Instruments USB6002) is used to monitor the count rate from the SPD and to stabilize the bias of MZI in real time.

The feasibility of proposed scheme on improving the dynamic range is verified and shown in Fig. 3. First, we take the MZI out from the experimental setups, i.e. PC-OTDR without external gates. After a 420 seconds measurement, we obtain a curve for the 70 m long fiber link with a connector located at 20 m from the beginning. The curve is in blue and given in Fig. 3(a). From this curve, a dynamic range of 19.0 dB at the end of the curve can be obtained. Then, we test the performance of externally time-gated PC-OTDR system. The MZI is acting as a high-speed OSW with an extinction ratio of 29.0 dB. The 50 ns gated signals are generated by the FPGA. In order to make a fair comparison, the total count rate in a 50 ns period is set as the same as the total count rate without gating operation, which is $4.69\times10^4$ Hz in our experiment. This is achieved by improving the power injected into the FUT. By adjusting the delay of the gated signal, we can scan and measure the OTDR trace for each segment and reconstruct the whole OTDR curve. The orange curve as shown in Fig. 3(a) is the recovered PC-OTDR trace with external gates. The fiber length of each segment is 5 m and 14 segments are obtained in the experiment. Each segment is obtained with a measurement time of 30 seconds corresponding to a total measurement time of 420 seconds - the time consumption for adjusting the delay of gates is negligible in our system. Again, from the orange curve, a dynamic range of 30.0 dB is observed in our experiment.

As shown in Fig. 3(a), the dynamic range for the case with external gates has a 11.0 dB improvement comparing with the case without gates. The improvement in dynamic range is very close to the ideal case, which should be 11.5 dB according to (5) (determined by the number of segments divided into, i.e. 14 segments in our demonstration) - the 0.5 dB decline is caused by the limited extinction ratio of MZI, i.e. 29.0 dB in our experiment. Furthermore, Fig. 3(b) and (c) are the results of that dynamic range change with different extinction ratios and gate widths, respectively. Firstly, we fix the gate width at 50 ns, and measure the improvement of the dynamic range with different extinction ratios in 70 m optical fiber. The change of extinction ratio is realized by varying the amplitude of the electronical gated signal feeding to MZI in our experiment. The theoretical prediction based on (4) is obtained, with $N_b \sim 7000$, $m \sim 500$. The result is the solid line shown in Fig. 3(b). It is obvious that the dynamic range increases with the increase of extinction ratio until it tends to be stable with extinction ratio greater than 25.0 dB. Then we fix the extinction ratio at the maximum value of 29.0 dB, and measure the improvement of the dynamic range with different gate widths changed from 5 ns to 400 ns. Similarly, we draw the theoretical prediction curve according to (5) as a comparison, for $N_b \sim 7000$, $E_R \sim 29$. The results are shown in Fig. 3(c). It is easy to obtain the inverse proportional function relationship between the change of dynamic range and gate width.

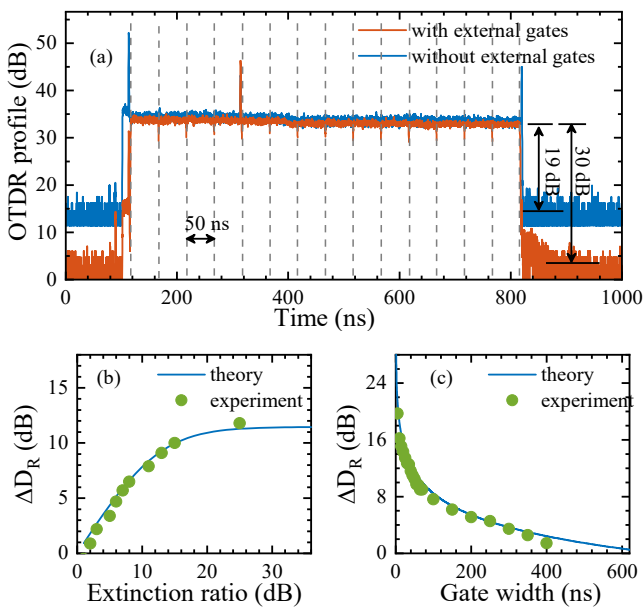

Fig. 3. Experimental results of proposed scheme. (a) OTDR traces of ~70 m fiber link obtained without (blue curve) and with external gates (orange curve). (b) and (c) are the variation of dynamic range with extinction ratios and gate widths, respectively.

With the improved dynamic range from the externally time-gated scheme, we are able to detect loss event which is drowned out by the noise in the case without gate. To verify, we perform a single gate test at the specified location. In this measurement, the FUT consists of two pieces of 20 m long multimode fibers. A micro-bend loss is introduced at around 32.5 m place along the FUT. To measure the micro-bend loss, a 50 ns gated signal is generated and delayed to cover it. The results are shown in Fig. 4. The blue curve is a reference curve indicating the result without external gate. Note that the residual three peaks out of the 50 ns gate is due to that the MZI employed in our experiment has a limited extinction ratio. The orange curve is the result for the case with external gate. In the two measurements, the accumulating time (60 seconds) and total count rate ($6.47 \times 10^3$ Hz) from the SPD are keeping the same. As shown in Fig. 4, it can be seen that a small loss event is observed in the case with external gate, while the curve is almost flat for the case without gate. To show it much clearer, we fit the two curves by using an adaptive filtering algorithm [18], as shown in the inset of Fig. 4. A loss event of 0.37 dB caused by the micro-bend is obtained due to the increase of dynamic range, while such a loss event is not observed in the case without gate. Note that the 0.37 dB loss is also observed by using a power meter to monitor the output port of the FUT. Hence, our external gated PC-OTDR is a very effective way to observe micro-bending loss along the fiber under test with high spatial resolution, i.e. less than 9 cm in our PC-OTDR testbed.

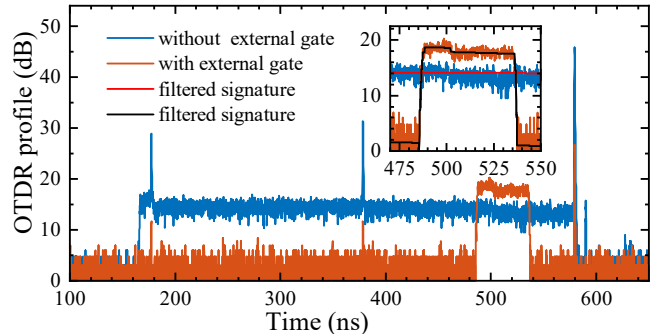

Fig. 4. Detection of micro bending by PC-OTDR with (orange curve) and without (blue curve) external gate. The red line and black line in the inset are the results after adaptive filtering for both cases of with and without external gating.

## IV. CONCLUSION

In this paper, an externally time-gated photon counting optical time-domain reflectometry based on high-speed optical switch has been proposed and experimentally demonstrated. Our externally time-gated scheme improves the dynamic range of PC-OTDR, which is limited due to the saturation count rates of single photon detectors caused by deadtime. We first carry out the theoretical analysis and simulation. It shows that the dynamic range of the externally time-gated system increases with the extinction ratio of the MZI, and decreases with the increase of the gate width - the one without gating operation corresponding to the lowest dynamic range. In our experimental investigation, the dynamic range of externally time-gated PC-OTDR is improved by 11.0 dB, which corresponds to dividing the whole OTDR curve into 14 segments by using external gating. More important, thanks to the improvement of the dynamic range with external gates, a micro-bend loss event with a loss of 0.37 dB has been observed in the experiment. Such a bend event has been previously submerged in the noise for the case without external time-gates.

One may think about that the dynamic range can be further improved by dividing the PC-OTDR curve into more numbers of segments by using our scheme. In theory this thought is true. While for implementation, we need to think about the speed of the optical switch and the pulse width of electronic gating signal. On one hand, by using cost-effective FPGA to generate electronic gating signal, the minimum gating width could be around 1 ns. On the other hand, the idea for the external gating scheme is that one can increase the photon counting number within the gating period by increase the power of probe signal. However, this maximum power is limited by the "pile-up" effect as discussed in Appendix B. Taking all the properties into account, our scheme has the characteristics of easy to implement and suitable for any PC-OTDR system. In addition, our results indicate that the proposed externally time-gated scheme can paves an avenue for achieving a PC-OTDR system with high dynamic range and has great potential in applications for high performance fiber test.

APPENDIX A: DEFECTS OF INTERNAL GATING OPERATION

In this appendix, we show the defects of the internal gating operation. The SPD (Excelitas Technologies, SPCM-AQRH) can work under so-called internally gated mode. One could think about developing a gated PC-OTDR with internally time-gated SPD, i.e. turning on the SPD at a certain period with an electronic gated signal and scanning the whole OTDR trace by moving the gate. We test this method in our experiment and the results are shown in Fig. 5(a). However, the gate only turns the output circuit of the SPD on or off, while it does not turn the Si-APD on or off, which means that the Si-APD still works under free running mode. Thus, the SPD keeps suffering from the saturation count rate under the internally gated mode. Furthermore, as shown in Fig. 5(a), a fake count event has been observed at the rising edge of the gate, which degrades the performance of PC-OTDR. It may be caused by the electrical (or photo) luminescence effect of the semiconductor material itself [21], i.e. the SPD operating under the internally gated mode may cause photon radiation phenomenon. To compare, Fig. 5(b) is the results of our scheme, in which the gated signal is applied to the OSW. Fig. 5 clearly shows the advantages of our proposed scheme: on one hand there is no fake peaks; on the other hand, the dynamic range is much higher than the case without gating operation.

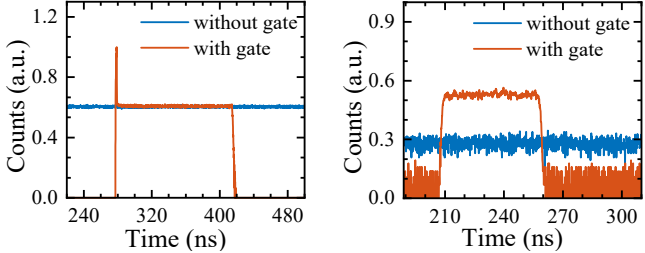

Fig. 5. The normalized results of SPD working under the (a) internally gated mode, (b) externally gated mode.

APPENDIX B: SIMULATION PLATFORM OF EXTERNALLY TIME-GATED PC-OTDR

Based on the theoretical model, an externally time-gated PC-OTDR simulation platform is established. It is mainly composed of three parts: Rayleigh scattering module [22], Fresnel reflection module [23] and single-photon detection module [24].

In our simulation, the dark count and dead time of the SPD are set as 200 Hz and 20 ns respectively, at the meanwhile the time jitter is not considered. In order to shorten the simulation time, the fiber length is set to 60 m. Fig. 6 shows the simulated PC-OTDR traces with different extinction ratios, while the width of the gates is kept the same as 15 ns. From Fig. 6(a)-(c), when the extinction ratio is 40.0 dB, 20.0 dB and 5.0 dB, the increases of dynamic range are 6.03 dB, 6.00 dB and 5.63 dB, respectively. This is in good agreement with (4) in the main text. We can conclude that with the decrease of extinction ratio, the suppression of the signal outside of the gate becomes smaller, and the proportion of photon counts contributed from the Fresnel reflections increases, thereby reducing the dynamic range.

One important feature for PC-OTDR system is that the maximum counts is determined by dead time of SPD. This means that for a given width of gate, the peak power of the probe signal increases with the extinction ratio of the OSW. We analyze this phenomenon with our theoretical model, and the results are shown in Fig. 6(d). In our simulation, the pulse width of the probe signal is 1 ns. It shows that the higher the extinction ratio of the OSW is, the greater the peak power of the probe signal is needed. When that the extinction ratio is larger than 30.0 dB, the peak power of the probe signal is almost unchanged. In conclusion, the larger the extinction ratio the higher the dynamic range is.

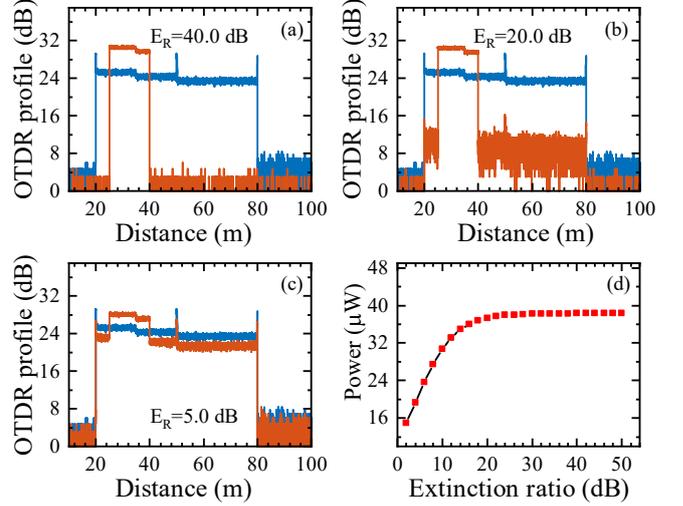

Fig. 6. Simulation results of PC-OTDR traces with extinction ratios of (a) 40.0 dB, (b) 20.0 dB, (c) 5.0 dB, respectively. The blue curve represents the result of PC-OTDR without gating operation. The result of externally time-gated scheme is represented by the orange curve using an optical switch with a gate width of 15 ns. (d) Peak power of probe signal varies with the extinction ratio.

Furthermore, the width of the external gate is an important parameter for PC-OTDR. In our theoretical analysis we keep the extinction ratio as 30.0 dB, and obtained the PC-OTDR results with different gate widths, as shown in Fig. 7. At the meanwhile the total count in each gate keeps the same, i.e. the maximum counts determined by the dead time of SPD, thus the peak power of the probe signal should be adjusted when changing the gate width. Assuming the power is $P_1$ and $P_2$, which correspond to gate widths of $W_{gate1}$ and $W_{gate2}$, respectively. Based on the (4) and (5) in the main text, we obtain straight forward

$$P_1 / P_2 = W_{gate2} / W_{gate1}. \qquad (6)$$

In general, if the gate width is increased by a factor $d$, the peak power of the probe signal is lowered by the same factor. As shown in Fig. 7(b) and (c), the wider the gate width, the smaller the pulse power required and the smaller the improvement of the dynamic range, which is in good agreement with (6) and (5).

Meanwhile, as shown in the inset of Fig. 7(a), the green curve has an anomaly segment with clear negative-slope, i.e. pile-up effect [25]. This effect arises because, in the SPD, only one event is detected at most (also related to the deadtime of the SPD), so that the photons appeared at the gate opening time have higher probability of being detected than those scattered at the back end. This effect is negligible when using the wide gate with low incident power, for example pulse widths of 100 ns and 150 ns in the simulation.

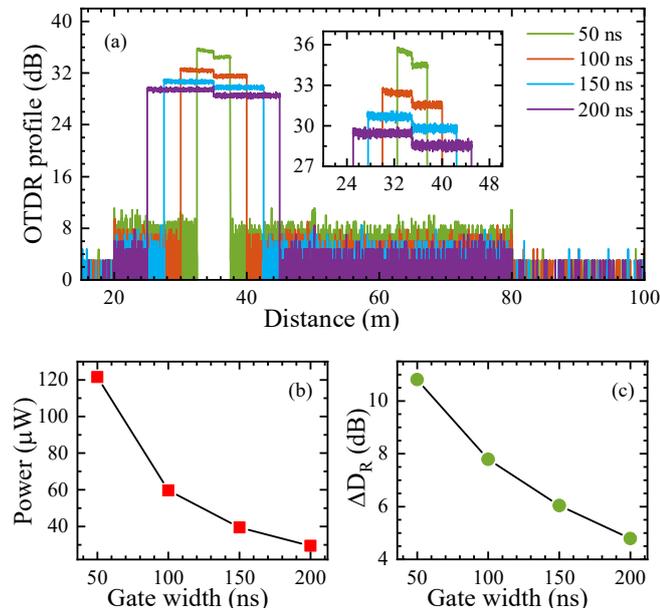

Fig. 7. The PC-OTDR simulation results corresponding to different external gate widths, when the extinction ratio of gated signal is 30.0 dB. (a) OTDR traces. (b) Peak power of probe signal with different gate widths. (c) The improvement of dynamic range with different gate widths for 60 m fiber under test.

We further study the pile-up effect theoretically, by using our theoretical model. First, by fixing the extinction ratio at 30 dB, we obtain a set of curves with different gate widths, and for each gate width the count rate from the SPD is the same. The results are shown in Fig. 8(a). Fig. 8(b) shows the slopes of the curves within the gate, obtained by piecewise-linear-fitting algorithm. Obviously, the slope is basically unchanged in the process of shortening the gate width from 100 ns to 60 ns. When the gate widths are 50 ns, 40 ns and 30 ns respectively, the slope of the initial segment decreases obviously and decreases with the decrease of the gate width. As the gate width is further reduced, the amount of change in slope becomes larger and larger, that is, the pile-up effect is more serious. This is due to the fact that in our scheme, the reduction in gate width means an increase in the power of the probe signal. Then we verify the change of the pile-up effect when only the peak power of the probe signal is changed. The results are shown in Fig. 8(c), at the meanwhile, the slopes of the curves within the gate are shown in Fig. 8(d), where the gate width and extinction ratio are fixed at 50 ns and 30 dB, respectively. It can be seen that the slope is unchanged at that the peak power equals to 0.02 mW. As the power increases to 1.0 mW, the slope changes suddenly and the slope of the initial segment becomes smaller. After that, the larger the peak power, the more significant of the slope gradient, which means that the pile-up effect is more obvious. In summary, in order to improve the dynamic range, we cannot improve the peak power or shorten the gating time unlimited, otherwise it will cause pile-up effect, resulting in test errors. In actual operation, we can set the gate width larger and adjust the power of the probe signal to make the count rate of SPD nearly saturated; then the optimal gate width can be obtained by continuously reducing the gate width and increasing the power until the obvious pile-up effect is about to appear in the test results. In this case, the reconstructed OTDR curve has the highest dynamic range.

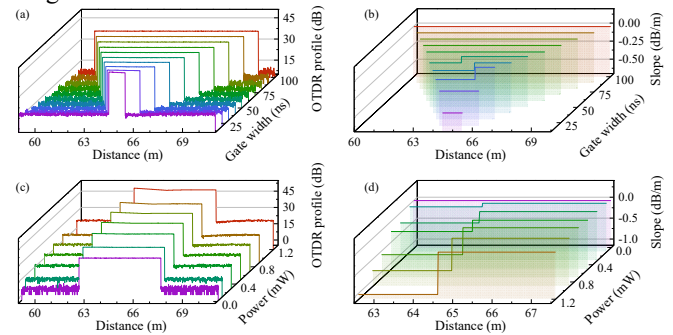

Fig. 8. (a) OTDR traces with external gate widths of 100 ns, 90 ns, 80 ns, 70 ns, 60 ns, 50 ns, 40 ns, 30 ns, 20 ns, 10 ns, respectively. (b) Slopes with different gate widths. (c) OTDR traces with different peak powers of probe signal, when the gate width and extinction ratio are fixed at 50 ns and 30 dB. (d) Slopes with different peak powers.